\documentclass[preprint,12pt, authoryear, nonatbib]{elsarticle}

\usepackage{amssymb}
\usepackage{amsmath}
\usepackage{url}
\usepackage{graphicx}
\usepackage{amssymb}
\usepackage{caption}
\usepackage{subfig}
\usepackage{slashbox}
\usepackage{array}
\usepackage{multirow}
\usepackage{color,array}
\usepackage{apacite}
\usepackage{natbib}

\journal{Expert Systems with Applications}

\begin{document}

\begin{frontmatter}

\title{Improved Generalizability of CNN Based Lane Detection in Challenging Weather Using Adaptive Preprocessing Parameter Tuning}

\author[uiuc]{I-Chen Sang\corref{cor1}} 
\cortext[cor1]{Corresponding author}
\ead{ichens2@illinois.edu}
\author[uiuc]{William R. Norris}
\ead{wrnorris@illinois.edu}

\affiliation[uiuc]{organization={Department of Industrial and Enterprise Systems Engineering},
            addressline={104 S. Mathews Ave.}, 
            city={Urbana},
            postcode={61801}, 
            state={IL},
            country={US}}

\begin{abstract}
Ensuring the robustness of lane detection is essential for the reliability of autonomous vehicles, particularly in diverse weather conditions. While numerous algorithms have been proposed, addressing challenges posed by varying weather remains an ongoing issue. Geometric-based lane detection methods, rooted in the inherent properties of road geometry, provide enhanced generalizability. However, these methods often require manual parameter-tuning to accommodate fluctuating illumination and weather conditions. Conversely, learning-based approaches, trained on pre-labeled datasets, excel in localizing intricate and curved lane configurations but grapple with the absence of diverse weather datasets.
\par 
This paper introduces a hybrid approach that merges the strengths of both methodologies. A novel adaptive preprocessing method is proposed in this work. Utilizing a fuzzy inference system (FIS), the algorithm dynamically adjusts parameters in geometric-based image processing functions and enhances adaptability to diverse weather conditions. This preprocessing algorithm is designed to seamlessly integrate with all learning-based lane detection models. When implemented with CNN-based models, the hybrid approach demonstrates commendable generalizability across weather and adaptability to complex lane configurations. Rigorous testing on datasets featuring challenging weather showcases the proposed method's significant improvements over existing models, underscoring its efficacy in addressing persistent challenges associated with lane detection in adverse weather.
\end{abstract}


\begin{highlights}
\item This work proposed an Expert System-based image pre-processing method
\item The preprocessing method adaptively tuned the parameters with a Fuzzy Inference System
\item The preprocessing method was integrated with multiple current CNN-based lane detection models and improved their robustness under rainy weather conditions
\end{highlights}


\end{frontmatter}

\section{Introduction}
\label{intro}
Advanced Driver Assistance Systems (ADAS) were introduced to vehicle systems in order to enhance efficiency and reduce the risks in driving. ADAS has been shown successful in reducing injury and deaths by sharing the workload of the drivers\citep{injuries}. Common functions in ADAS are Lane Keeping Assistance systems (LKA), Adaptive Cruise Control (ACC), and Automatic Emergency Braking (AEB). 
\par 
While ACC and AEB operate with distance estimation in the front, LKA tackles information that includes complicated road conditions and diversity of lane patterns. As a result, the lane detection function is a widely studied technology\citep{ashape}. It is crucial for the research community to develop a lane detection algorithm that has high accuracy and adaptability for various weather conditions\citep{lanetrapezoid}.
\par 
Given the color and shape information in typical lines on the road, cameras are the most commonly used option in the lane detection task. Several key features are commonly used in visual-based lane detection. For instance, the geometric attributes of lanes are often employed to eliminate erroneous candidates, utilizing the parallel nature of lane lines as a guiding principle\citep{geometric_constraint}. On the other hand, color\citep{yuv} and edge\citep{lanetrapezoid} features have been widely applied in lane detection algorithms because of the visual dissimilarity between the lines and asphalt. Studies have also shown that the fusion between LiDAR and cameras can improve detection results compared to a single sensor.\citep{laser_camera}\citep{laser}
\par 
In the field of lane detection, convolutional neural networks (CNNs) have been the most commonly used approaches. CNNs are able to effectively learn through datasets and perform accordingly, making these models competitive in identifying the lane marks in images\citep{lane_review_new_2}. However, cameras are sensitive to weather changes. Specifically, when the road is covered with water on a rainy day, the appearance of the lines and asphalt becomes similar. The flowing raindrops on the windshield, as well as the fast-moving wipers, pose extra difficulties for visual sensors\citep{lidar_rain}\citep{image_lidar_improve}. As a result, modifications need to be made to address the challenges posed by weather conditions, providing better inputs for CNNs, thereby enhancing the accuracy of the output.
\par 
In this article, a preprocessing algorithm that is compatible with state-of-the-art lane detection frameworks is proposed. The algorithm consists of geometric-based modules with tunable parameters in order to adapt to different weather conditions. Two closed-loop feedback fuzzy logic modules are leveraged to tune the threshold values. With the noise-tolerant nature of fuzzy inference systems (FIS), the algorithm is able to overcome challenging weather conditions and provide more consistent inputs for learning-based lane detection convolutional neural networks (CNNs). With this innovative combination of geometric and learning-based methods, an improvement in the detection accuracy across multiple CNNs and datasets is presented, showing the effectiveness and robustness of the proposed algorithm.
\par 
This article is organized as follows: Section II delves into previous research and the challenges encountered in the field of lane detection. In Section III, a comprehensive overview of the preprocessing algorithm, datasets employed, and the validation process is provided. Furthermore, Section IV encompasses the presentation of experimental results and in-depth discussions. Finally, in Section V, the article concludes by outlining the findings of this approach of this approach and discussing potential avenues for future research.

\section{Related Work}
Numerous works have been done in the lane detection field. These algorithms can be divided into geometric-based methods and learning-based methods. 
\subsection{Geometric-based methods}
Geometric-based methods, also known as traditional approaches, utilize geometric features to identify the lines. For example, authors of \citep{yuv} developed a method that was able to identify lane lines of different colors. The algorithm proposed in \citep{ashape} adopted a clustering approach that correlated pixels with similar color and brightness. Other than color information, the edges of lines were leveraged in \citep{lanetrapezoid}. Thanks to the clear pattern in most road configurations, the geometric-based methods showed good generalizability and decent accuracy in the aforementioned works. Additionally, pre-trained models are not needed in these approaches.
\par 
Despite the strengths mentioned in the previous paragraph, challenges in several aspects still exist for traditional approaches. Geometric-based methods highly depend on brightness and color information and therefore require hand-tuning parameters to adapt to various conditions. For example, the algorithm proposed in \citep{rain_threshold} showed a high accuracy of lane detection in rainy weather. However, after validation, it was found that the algorithm had difficulty working in lower illumination conditions unless the threshold was tuned manually. A previous work published by the authors \citep{fuzzy_traditional} proposed a geometric-based method that incorporated Fuzzy Inference Systems to adaptively tune the parameters. The proposed method in \citep{fuzzy_traditional} was shown to be effective across various weather conditions without the need for manual parameter-tuning. However, due to the complexity of curved lanes, many geometric-based studies are scoped on single, straight lanes \citep{ashape}\citep{lanetrapezoid}\citep{rain_threshold}\citep{fuzzy_traditional}.
\par

\subsection{Learning-based methods}
On the contrary, learning-based methods utilize pre-collected and labeled data to attain precise detection. Throughout the training phase, these models deduce a set of parameters that yield optimal detection outcomes. Consequently, unlike traditional approaches, learning-based methods eliminate the need for manual parameter adjustments. Furthermore, learning-based methods take advantage of deep layers that process data at higher dimensions, providing benefits for image segmentation. As a result, learning-based models enable the handling of curved and multiple lines with greater efficacy.\citep{lane_review_new}
\par 
Extensive learning-based algorithms have been published in the lane detection field. For instance, CLRNet\citep{clrnet} and \citep{clrernet} adopted a cross-layer approach to cover multi-scale features in the images, thereby achieving a high detection accuracy. RESA\citep{resa} dealt with the task with global features and performed pixel-wise identification. Different from standalone CNN models, authors of \citep{enet_sad} introduced the Self-Attention Distillation (SAD) model to improve the performance of existing models, such as ResNet\citep{resnet} and ENet\citep{enet}. All of the algorithms mentioned above showed successful detection results in multiple curved lanes.
\par 
Despite the adaptability to complicated line patterns, the difficulty faced in challenging weather conditions is seldom addressed in learning-based methods because of the considerable effort needed in data collection. An abundant lane detection dataset that includes various challenging weather conditions is needed to enhance the generalizability of the algorithms.  However, many commonly used datasets, such as TuSimple\citep{tusimple}, Caltech\citep{caltech}, and KITTI\citep{kitti} were only collected in clear daylight. Efforts have been made to enhance the diversity of scenes in various datasets. For instance, the CULane dataset\citep{culane} introduced several weather conditions, encompassing nighttime and foggy scenarios. However, it notably lacked challenging images depicting rainy and snowy conditions. In contrast, the DSDLDE dataset\citep{dsdlde2024}showcased a collection of 50 videos captured across a spectrum of weather conditions, spanning clear, rainy, and snowy weather. Regrettably, the official source did not include ground truth data.
\par 
In order to resolve the difficulty in data collection, works were proposed to raise the generalizability of learning-based methods. Authors of \citep{trainday1}\citep{trainday2}\citep{trainday3} trained the models using the daytime data and applied the neural networks on the nighttime test sets. However, this approach has not been extended to rainy and snowy conditions yet. The other approach that deals with the limitation of datasets is data generation. Studies proposed in \citep{transfernight} and \citep{foggy} generated nighttime and foggy images from clear daytime videos to train the lane detection networks. This method can significantly increase the variety of datasets and strengthen the robustness of learning-based methods. Nevertheless, the generative algorithms have not been applied in rainy and snowy weather conditions.

\subsection{Hybrid methods}

A hybrid approach is promising for combining the advantages of geometric-based and learning-based methods while addressing their limits. Hybrid image processing algorithms have been applied to medical images. The work presented in \citep{combine_medical} adopted both traditional quantitative image features and deep features that were extracted using pre-trained neural networks in a lung tumor detection task. The result showed that the accuracy was significantly improved compared to using only traditional features and deep features. In addition, hybrid approaches were also leveraged in texture classification tasks. Authors of \citep{hybrid_texture} applied the concept of traditional histogram calculation on feature maps to obtain the features' statistical information. Introducing this ``histogram layer'' presented a more accurate classification result. 
\par 

Similar attempts were also proposed in the lane detection field. The algorithm proposed in \citep{lane_combine1} incorporated traditional image preprocessing functions, such as range of interest (ROI) selection and inverse perspective transformation, with neural networks to conduct lane detection tasks. Moreover, the authors of \citep{lane_combine2} integrated the Adaptive Neuro-Fuzzy Inference System (ANFIS) with traditional geometric-based lane detection algorithms to achieve a better result. Both hybrid methods successfully combined the advantages of geometric-based and learning-based methods. 
\par 
However, the potential of the hybrid methods applied in lane detection algorithms' adaptability to weather conditions has not been explored yet. The work proposed in \citep{lane_combine1} incorporated Range of Interest (ROI) selection, figure interpolation, and perspective transformation before applying neural networks to the images. The parameters used in the aforementioned traditional image processing modules were not adaptable to various environmental conditions. The work proposed in \citep{lane_combine2} though combining horizontal gradient calculations with an ANFIS, was only applied to previously cropped images where only the asphalt area was included. In addition, \citep{lane_combine2} required an initial estimate of the lane marks, which potentially limited the application of the method.
\par
As a result, it was believed that hybrid methods have great potential in coordinating the advantages of both geometric-based and learning-based methods the potential of hybrid methods is yet to be explored. This article presents a geometric-based adaptive preprocessing algorithm. When integrated with CNN-based lane detection models, the proposed function significantly improved multiple CNN models' robustness over challenging weather conditions.
\section{Methodology}
\begin{figure*}[b!]
\centering
\includegraphics[width=0.9\textwidth]{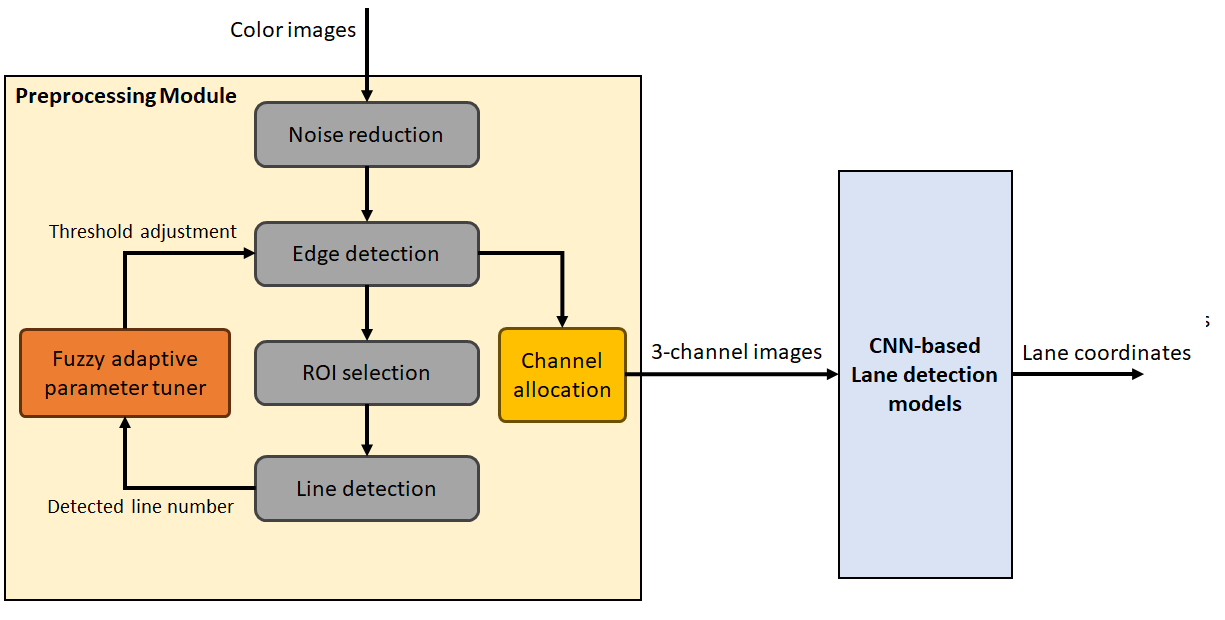}
\caption{The flow chart for the overall structure is shown in this figure. The proposed preprocessing framework was connected with CNN-based lane detection algorithms}
\label{fig:flowchart}
\end{figure*}
This section presents the proposed preprocessing framework and all included functions in detail. In order to integrate with most lane detection models, the framework processed color images and output images with the same size and number of channels. 
\subsection{Algorithm description}
The overall structure of the algorithm is shown in Figure \ref{fig:flowchart}. The input images were processed by the Noise reduction module, the Edge detection module, the ROI selection module, and the Line detection module sequentially. The threshold parameter in the Edge detection module was adjusted by a Fuzzy parameter tuner in each iteration. That is, the measurement acquired in frame $N$ was processed by the parameter tuner which determined the parameter used in frame $N+1$. The functionality and the details of each module are described in the paragraphs below.

\begin{itemize}
    \item Noise reduction
\end{itemize}

To simplify the calculation, the images were transformed to grayscale when imported into the preprocessing module. A bilateral filter\citep{bilateral} was leveraged to suppress the noise in the images, enhancing the quality of the edge detectors. The bilateral filter calculated the weighted average of neighboring pixels with the weights determined by assigned spatial and color variance values. The mathematical expression of the bilateral filter was 
\begin{equation}
    K_{x_i}(x) = \frac{1}{W}exp\left( \frac{||x_i-x||^2}{\sigma_s}\right )exp\left ( \frac{|I(x_i)-I(x)|^2}{\sigma_I} \right)
\end{equation}
where $x$ was the center pixel and $x_i$ was the neighboring pixels in the kernel. $\sigma_s$ and $\sigma_I$ represented the pre-assigned variance value in space and intensity respectively, and $W$ is the normalization parameter. A kernel size of $7\times7$, spatial variance of 50, and intensity variance of 25 were used in this work. 
\par
\begin{figure}[t]
\centering
\subfloat[]{\includegraphics[width=3in]{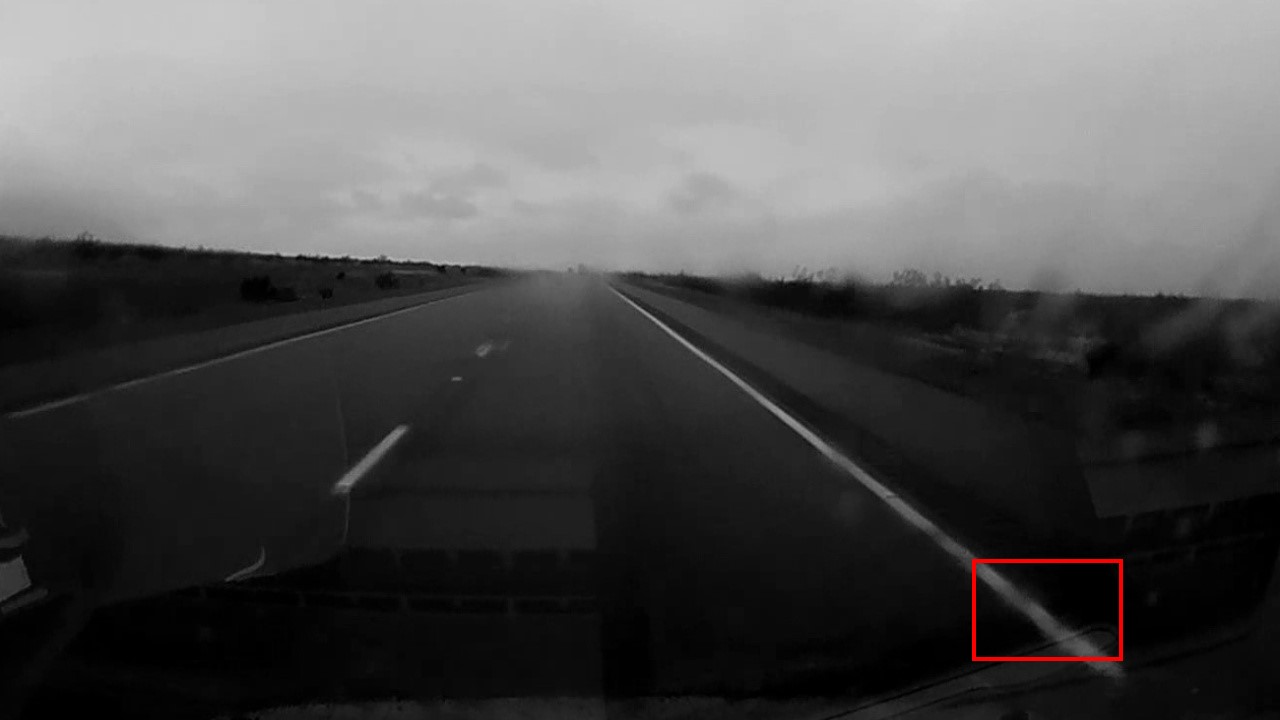}}
\\
\subfloat[]{\includegraphics[width=1.4in]{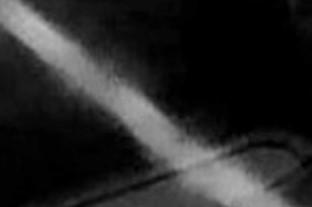}}
\hfil
\subfloat[]{\includegraphics[width=1.4in]
{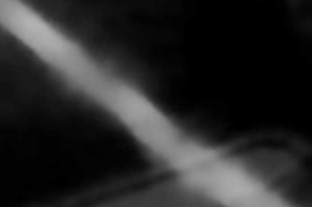}}
\\
\subfloat[]{\includegraphics[width=1.4in]
{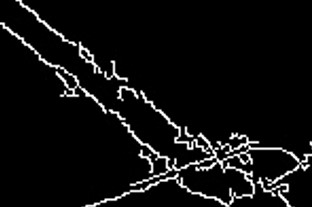}}
\hfil
\subfloat[]{\includegraphics[width=1.4in]
{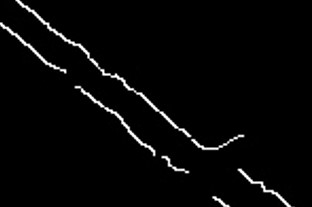}}

\caption{An example of the effect of the bilateral filter. Figure (b) and (c) show the magnified view of the red box in (a) before and after bilateral filtering. Figures (d) and (e) show the edge detection results of (b) and (c), respectively. }
\label{bilateral_effect}
\end{figure}

 
The contribution of the noise reduction module is shown in Figure \ref{bilateral_effect}. Figure \ref{bilateral_effect}(b) and (c) show the magnified image inside the red box before and after the filtering process, while (d) and (e) show the edge detection result. It is clear that the noise reduction module made a substantial improvement in the quality of the edge detection results. The erroneous detections resulting from image noise were effectively eliminated.

\begin{itemize}
    \item Edge detection
\end{itemize}
After reducing the noise in images, a canny edge detection function\citep{canny} was utilized in this algorithm. The Canny edge detection function leveraged a $3\times3$ filter to detect the gradient of the images in the x and y directions. The L2 summation of both gradient values then went through the non-maximum suppression before the thresholding process. 
\par 
There were two threshold parameters required in the Canny edge detection function, $Th_{high}$ and $Th_{low}$. The pixels with gradient values higher than $Th_{high}$ were reserved while the ones lower than $Th_{low}$ were deleted. On the other hand, the pixels with gradient values between $Th_{high}$ and $Th_{low}$ were only reserved when neighboring any candidates that were greater than $Th_{high}$. As a result, a properly tuned threshold value is pivotal for the edge detection result. Typically, the threshold values are manually tuned with $Th_{low}$ set to 1/3 of $Th_{high}$ \citep{canny}. In this work, $Th_{high}$ was adjusted by the fuzzy parameter tuner, and $Th_{low}$ was 1/3 of $Th_{high}$.

\begin{itemize}
    \item ROI selection
\end{itemize}
The number of candidate pixels hugely influenced the processing time of the preprocessing framework. In order to select better candidates among the edge pixels derived by the Canny edge detector, an ROI selection module was introduced. 
\par
\begin{figure}[t]
\centering
\subfloat[]{\includegraphics[width=2.4in]{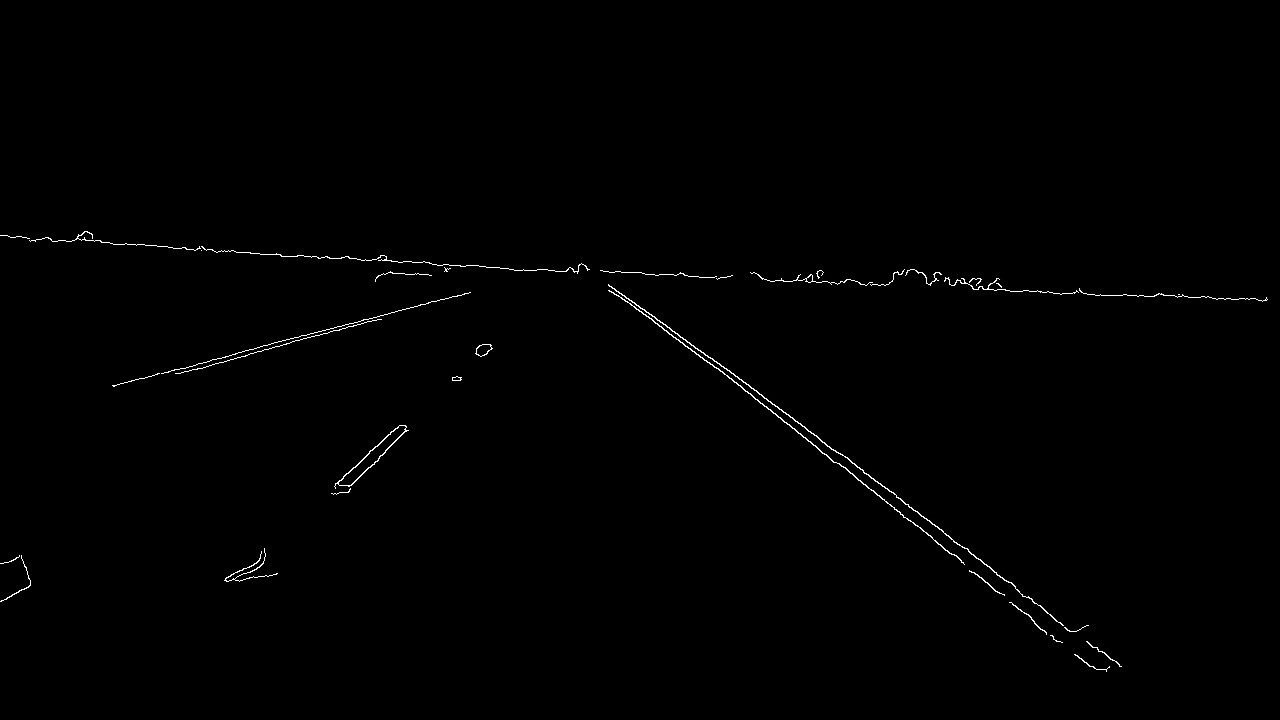}}
\hfil
\subfloat[]{\includegraphics[width=2.4in]{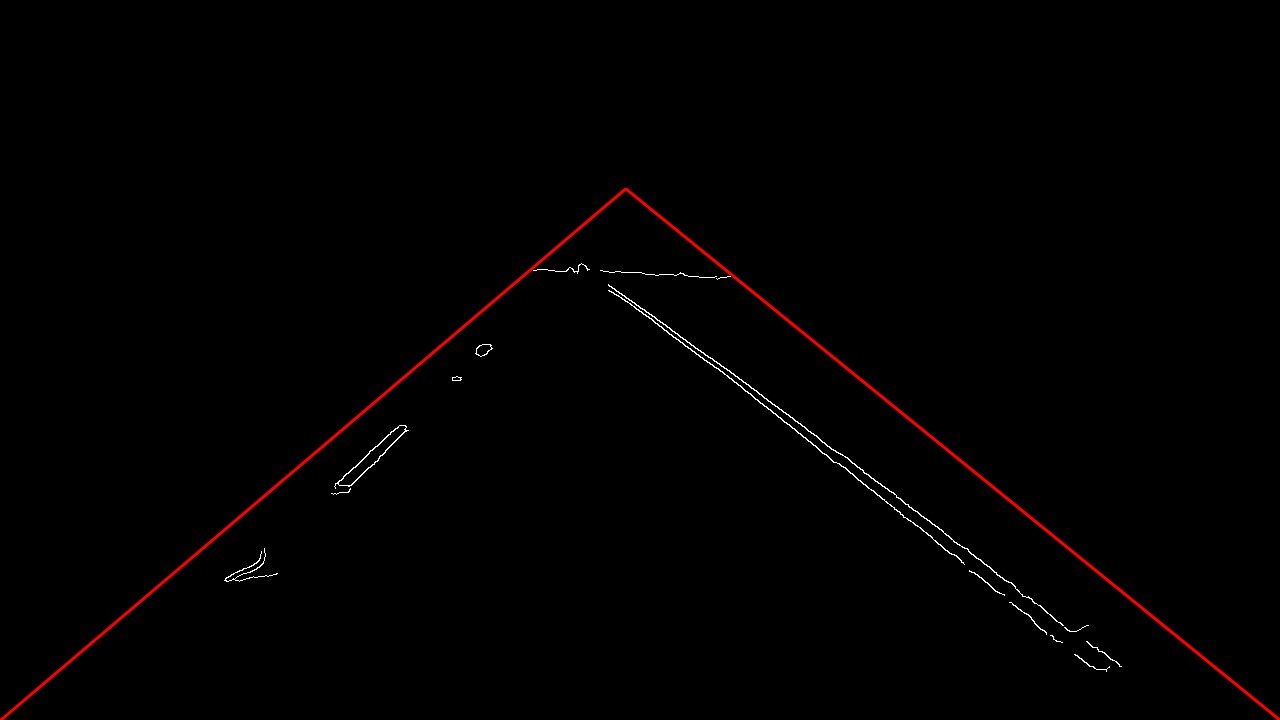}}
\caption{An example of the detection result of a frame before (a) and after (b) applying the triangular ROI.}
\label{roi_effect}
\end{figure}

 
Typical roads consist of parallel lines that converge to a vanishing point in images\citep{geometric_constraint}. Therefore, with a triangular ROI, the possibility of capturing at least one lane in the image is high. When the slope is limited, the vanishing point resides around the center of the images. In this study, an ROI region that covers the vanishing point with a margin was adopted. Figure. \ref{roi_effect} shows the edge detection result before (a) and after (b) the ROI selection process. The apex of the ROI was set at the center of the image with 3/4 of the image height.

\begin{itemize}
    \item Line detection
\end{itemize}
A line detection function was incorporated into the process to tune the threshold of the Canny edge detection module. It was worth noting that the line detection result did not directly assist the lane detection process. Instead, it was leveraged as a measure to evaluate the performance of the current threshold value. The reasoning behind the design is that learning-based methods are more efficient and accurate when plotting curved lanes, while geometric-based line detectors provide tuning references for the parameter tuner more effectively. 
\par 
Considering the design of the triangular ROI and the typical structure of roads, it was expected that the ROI would contain 1-2 lanes, which ranged from 1-3 line marks. As a result, when applying line detection modules, the detected line number should fall into a fixed range if the edge detection module is correctly tuned. To be more specific, when the edge detection module outputs more edge pixels than necessary, a higher detected line number is found; therefore, an increase in the threshold value is needed in this case. 
\par 
The commonly used Hough Transform \citep{hough} was adopted to identify and quantify the line numbers in the images. The Hough Transform method generates straight lines in all directions, with each line centered on every edge pixel present in the images. A ``vote'' value is assigned to each line by tallying the number of edge pixels that align with the generated line. Lines with higher vote counts exhibit a stronger connection with edge pixels, thus making them preferred candidates for representing lane lines.
\par 
An angular resolution of 1 degree and a minimum intersection of 3 was adopted in this work. The detected line count was then used to tune the threshold value in the Canny edge detector, moderating the number of candidate pixels. The proposed closed-loop feedback process maintained a proper number of candidates, thereby eliminating the excessive processing time of Hough Transform.

\newpage

\begin{itemize}
    \item Fuzzy parameter tuner
\end{itemize}

An innovative parameter-tuning method is proposed in this work. Inspired by control algorithms where only ``steering'' decisions are output, the parameters were tuned relatively. To be more specific, the parameter tuner decided the direction (increase/decrease) and the level (more/less) of the applied change instead of outputting an absolute threshold value. 
\par 
As mentioned previously, given the road structure and the inclusion of the ROI, the number of detected lines in the triangular ROI should fall into a fixed range. Therefore, the detected line number was leveraged as the ``observable parameter'' that could be used to tune the threshold value of Canny edge detector. 
\par 
A fuzzy inference system (FIS) was introduced in this process. Among the types of FIS, Sugeno FIS allows more flexibility in output design and optimization, while Mamdani FIS is more intuitive in translating expert's knowledge. In this paper, the Mamdani system was chosen for its emphasis on simulating the human operator's decision during parameter tuning and its mathematical simplicity, which avoids the complex optimization functions required by the Sugeno system. FIS is based on rulebases that reflect human decisions. Therefore, fuzzy inference systems are also referred to as expert systems. Similar to human cognition processes, the input signals are fuzzified and labeled with linguistic variables before being processed by the rulebase. The linguistically labeled decisions made by the system are then defuzzified and projected to a numerical output value. 
\par 
To be more specific, after the input values were categorized using linguistic parameters (such as ``Too few'', ``Few''...), the rulebase was applied to reflect the logic human operators used to tune the threshold. The rulebase listed all input conditions and the corresponding linguistic outputs (such as ``Add'', ``Minus''...). A defuzzification process that projects the linguistic output into numerical output was implemented accordingly.
\par 
The fuzziness of FIS increases the algorithm's robustness when dealing with noises and the varying environment. In addition, the combination of linguistic variables and rulebases enables the users to design complicated projection functions intuitively. 
\par 
\begin{table}[t]
\renewcommand{\arraystretch}{1}
\caption{The rule base of the fuzzy edge detection module. }
\label{rule_base}
\centering
\begin{tabular}{|c|c|}
\hline
Detected line number & Action to Canny threshold \\
\hline
Too few & Minus2\\
\hline
Few & Minus1\\
\hline
Good & Zero\\
\hline
Many & Add1\\
\hline
Too many & Add2\\
\hline
\end{tabular}
\end{table}

\begin{figure}[t]
\centering
\subfloat[]{\includegraphics[width=3.2in]{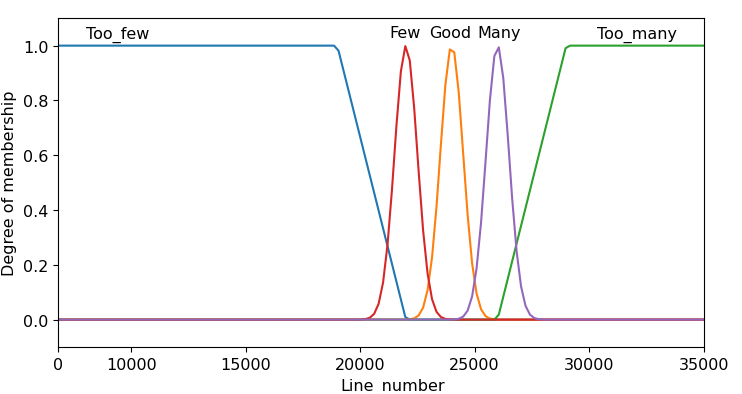}}
\hfil
\subfloat[]{\includegraphics[width=3.2in]{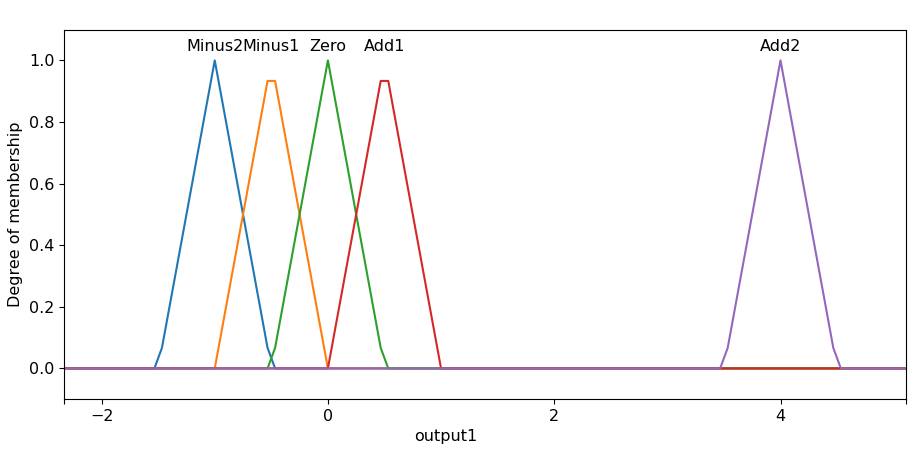}}
\caption{The membership functions of the (a) input and (b) output of the threshold adjusting module. }
\label{fuzzy}
\end{figure}

A single input single output (SISO) FIS system was leveraged in the parameter tuning process. Five input membership functions (Fig. \ref{fuzzy}(a)) were defined according to the detected line count. Based on the membership functions, the probability of the input belonging to each linguistic category was calculated. The rulebase used in this work is listed in Table \ref{rule_base}. When the line number detected was fewer than the expected value, the threshold of the Canny edge detector was reduced in order to introduce more information to the system and vice versa. The decisions made by the FIS were then defuzzified with the output membership functions(Fig. \ref{fuzzy}(b)). 
\par 
All membership functions were designed experimentally. The initial Canny threshold value was set to 1 in this work because it allows the most information from the images to be processed. The observed threshold variation ranged from 20 to 40 in one scene. Assuming the frame rate to be 30fps, the increment was designed to be $\pm1.5$/frame in order to converge within 30 frames. Five fuzzy inference rules were designed in the proposed method. Case ``Zero'' adjusted the threshold value for $-0.5\sim0.5$. The cases ``Add/Minus 1'' tuned the threshold value for $\pm0\sim0.5$. The case ``Minus 2'' adjusted threshold value for $\pm0.5\sim1.5$. However, to raise the threshold to the appropriate value effectively and thereby enhance the detection rate, an asymmetric design was adopted for the ``Add2'' function. This off-center function raised the threshold value quickly when the experiment was initiated, helping the system to stabilize within fewer frames. Since an initial threshold of 1 was selected, the adjustment value of ``Add 2'' was designed to be $+3.5\sim4.5$ in order to accelerate the convergence process.
\par 
With the fuzzy parameter tuner, the system controlled the number of edge pixels that needed processing, enhancing the processing speed. A detailed ablation study was performed in \citep{fuzzy_traditional} to show the capability of the proposed tuner.

\begin{itemize}
    \item Channel allocation
\end{itemize}

\begin{figure}[!t]
\centering
\includegraphics[width=3in]{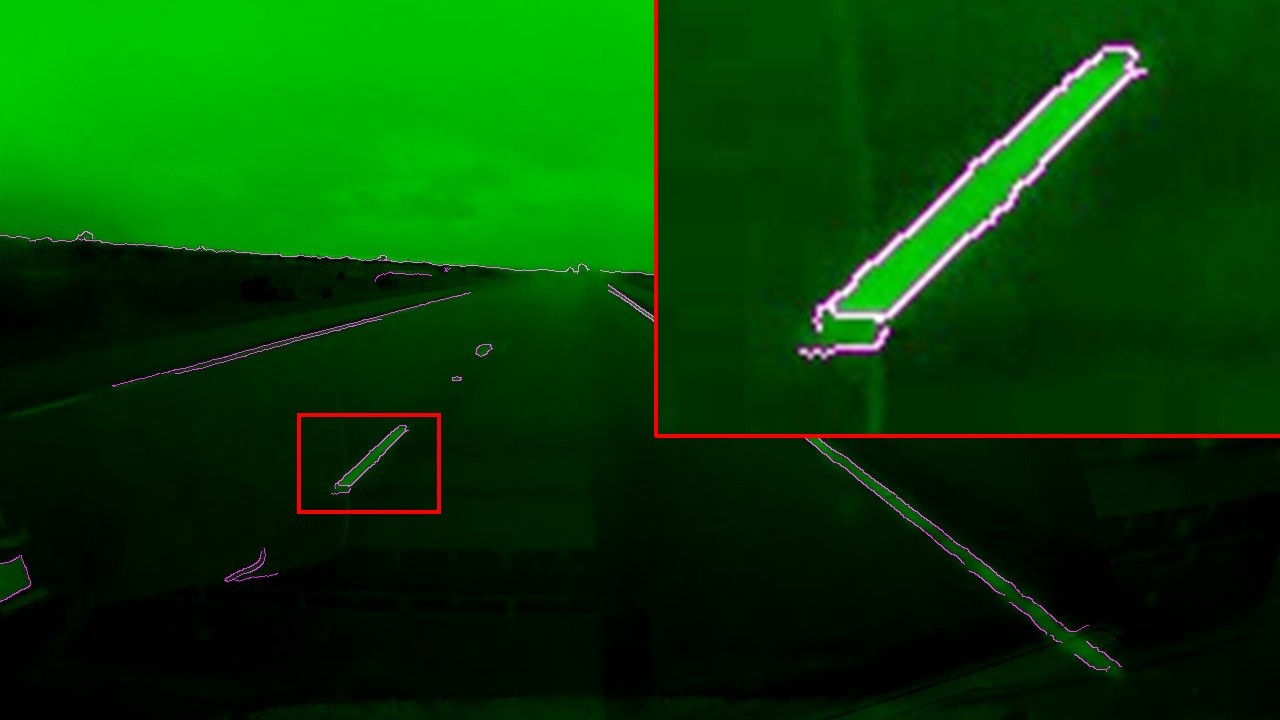}
\caption{The result of the channel allocation process of the proposed method.}
\label{channel}
\end{figure}

At the end of the preprocessing algorithm, a channel allocation process was applied. Due to the adaptive parameter tuning capability of the algorithm, the edge detection result was able to outline the lane lines effectively. Therefore, the edge detection results were included in the final output to the CNN-based lane detection models.
\par 
Based on the experiments conducted, the green channel in the images carried more information on the white and yellow lines. Therefore, in the channel allocation process, the detected edge images were normalized to 8-bit and used to replace the red and blue channels of the original images. An example of the allocated result is shown in Figure. \ref{channel}. The figure shows that the green channel preserves most image features of the lines while the edge channels well-outlined the marks. This approach maintained the dimensions and the number of channels of the input image, enabling seamless integration with most CNN-based lane detection models.

\begin{table}[t]
\caption{The key parameters adopted in the preprocessing module.}
\centering
\begin{tabular}{|c|c|}
\hline
Parameters                                                                              & Values                                                         \\ \hline
ROI  apex                                                                               & (3/4*Height, 1/2*Width)                                        \\ \hline
Bilateral filter kernel                                                                 & 7                                                              \\ \hline
\begin{tabular}[c]{@{}c@{}}Bilateral filter variance\\ (spatial/color)\end{tabular}     & 25/50                                                          \\ \hline
Initial Canny threshold                                                                 & 1                                                              \\ \hline+
\begin{tabular}[c]{@{}c@{}}Hough Transform resolution\\ (distance/angular)\end{tabular} & \begin{tabular}[c]{@{}c@{}}1 (pixel)\\ 1 (degree)\end{tabular} \\ \hline
\end{tabular}
\end{table}

\section{Experiment}
To show the performance and the adaptability of the proposed preprocessing algorithm, the algorithm was incorporated with three different CNN-based lane detection models and tested on datasets with clear and rainy weather conditions. The datasets used to validate the proposed algorithm were collected in the real world under clear weather conditions (Tusimple dataset) and rainy weather conditions (DSDLDE dataset). The details of the experiments are presented in detail in the following subsections. 
\subsection{Experiment setup}
The experiment was designed to simulate the on-road driving conditions where all incoming images were processed sequentially. The validation experiments were performed using a PC with an Intel core i9, 3.0-5.8 GHz CPU with 64GB RAM, and a Nvidia RTX A5500, 24GB.

\subsection{Datasets}
\begin{itemize}
    \item Tusimple\citep{tusimple}
\end{itemize}
The Tusimple dataset is a commonly used testbench in the lane detection field. It contains 6766 clips of 20 frames, with the last frame being labeled. Specifically, the training set encompasses 3626 clips, the validation set comprises 358 clips, and the test set encompasses 2782 clips. Notably, all images in the dataset are in RGB color format and exhibit a resolution of $1280\times 720$.
\par 
Despite the uniform collection location being the highways of the United States, the Tusimple dataset exhibits a diverse array of line colors, configurations, and quantities. The dataset mostly captures bright and clear scenarios, with variations in the angles from which sunlight illuminates the scenes.

\begin{itemize}
    \item DSDLDE\citep{dsdlde2024}
\end{itemize}
The DSDLDE dataset specializes in the diverse weather conditions included. DSDLDE consists of 50 video clips, where 27 were collected during daytime and 23 were captured during nighttime. The clips ranged from 900 to 2800 frames. In each category, three weather conditions (clear/rainy/snowy) were included. All images were at the resolution of $1920\times 1080$ at RGB color space. 
\par 
Unlike the Tusimple dataset, the DSDLDE dataset was collected in both the United States and South Korea. A variety of scenes, including rural, city, caves, and highway, were covered. The dataset was constructed by multiple vehicles and camera setups. Therefore, different fields of view (FOV) were found in the dataset. For example, the area occupied by the hood or the interior of the vehicles ranged from 0 to roughly 40\% of the images. Since this work is focused on algorithms' adaptability across various weather conditions, efforts were made to simplify the comparison and focus on the weather variable. The images were cropped to an FOV that is similar to that of the Tusimple dataset and resized to $1280\times 720$. 
\par 
On the other hand, the biggest drawback of the dataset is the lack of labels and evaluation metrics. Therefore, to show the quantitative detection result of the DSDLDE dataset, a labeling GUI was made by the authors. The GUI assists the users in labeling any lane detection datasets and outputs the ground truths in the format of the Tusimple dataset, enabling quantitative comparison. The labeling GUI can be found at \url{https://github.com/ichensang/Lane-labeling}. The labeled results were uploaded to \url{https://github.com/ichensang/DSDLDE-labels}. 

\subsection{Training and testing process}
The proposed preprocessing algorithm was integrated with three different CNN-based lane detection models and trained respectively. The models used in the experiment were CLRNet-Resnet18\citep{clrnet}, RESA\citep{resa}, and Enet-SAD\citep{enet_sad}. Since the DSDLDE dataset did not provide a training set, all models were only trained using the training set of the Tusimple dataset. 
\par 
The initial setting of the parameters in the preprocessing algorithm was already covered in the Methodology section. On the other hand, most parameter settings for the lane detection models followed the original codes. Only the batch sizes were adjusted due to hardware limitations. The batch sizes of CLRNet-Resnet18, RESA, and Enet-SAD are 10, 4, and 5. A  comprehensive summary of the key parameters is shown in Table \ref{parameter_NN}.

\begin{table*}[t]
\centering
\caption{The parameters adopted in each neural network.}
\resizebox{\textwidth}{!}{%
\begin{tabular}{|c|c|c|c|c|c|c|}
\hline
Network  & Epochs & Batch Size & Optimizer & Learning Rate & Backbone & GPU number \\ \hline
CLRNet   & 70     & 10         & AdamW     & 1e-3          & Resnet18 & 1          \\ \hline
Resa     & 300    & 4          & SGD       & 2e-2          & Resnet34 & 1          \\ \hline
Enet-SAD & 150    & 5         & SGD       & 1e-2          & Enet     & 1          \\ \hline
\end{tabular}%
}
\label{parameter_NN}
\end{table*}

\par 
In order to simulate the scenario of an onboard driving task where the parameter tuner adjusts the threshold values with time, the testing process was revised accordingly. The test images were processed one at a time following sequential order. For the Tusimple dataset, the images in the test set were not continuous frames. Therefore, the images were processed sequentially according to the file names. The DSDLDE dataset, however, contained consecutive frames from multiple clips. So the test of the DSDLDE video clips was executed on a per-clip basis with the images processed sequentially with the timestamp.
\par
\begin{figure}[t]
\centering
\subfloat[]{\includegraphics[width=2in]{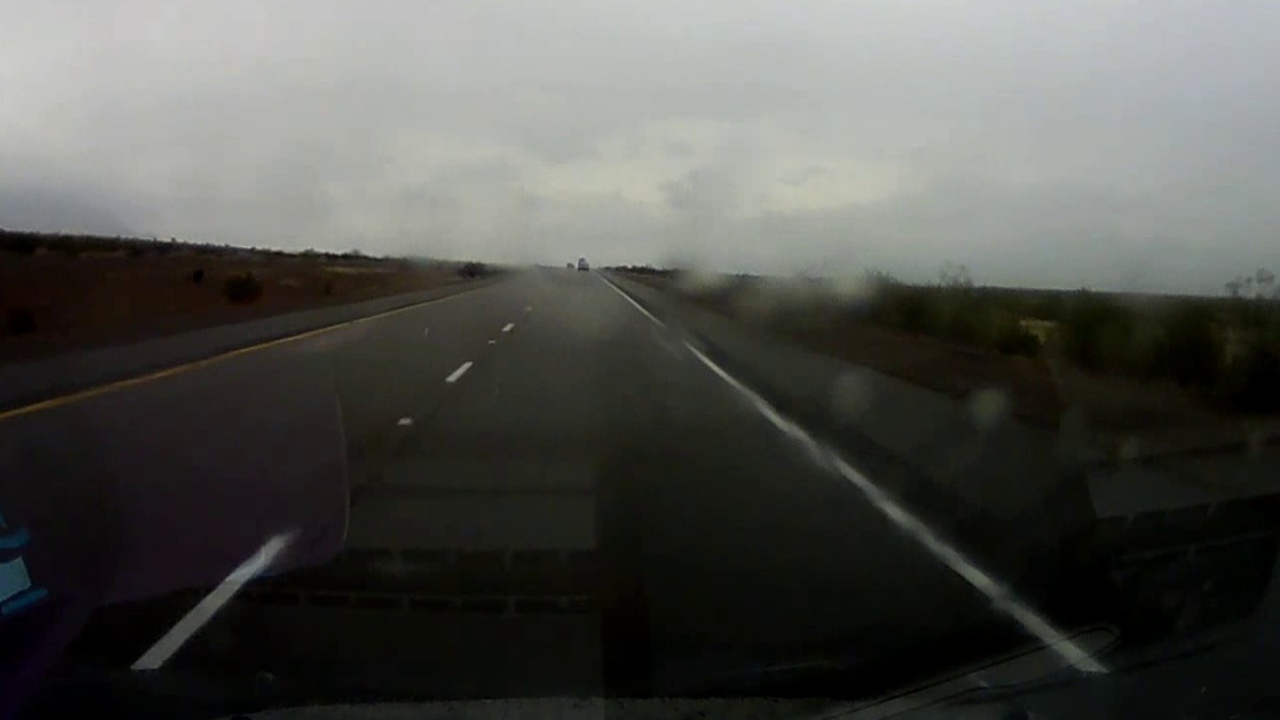}}
\hfil
\subfloat[]{\includegraphics[width=2in]{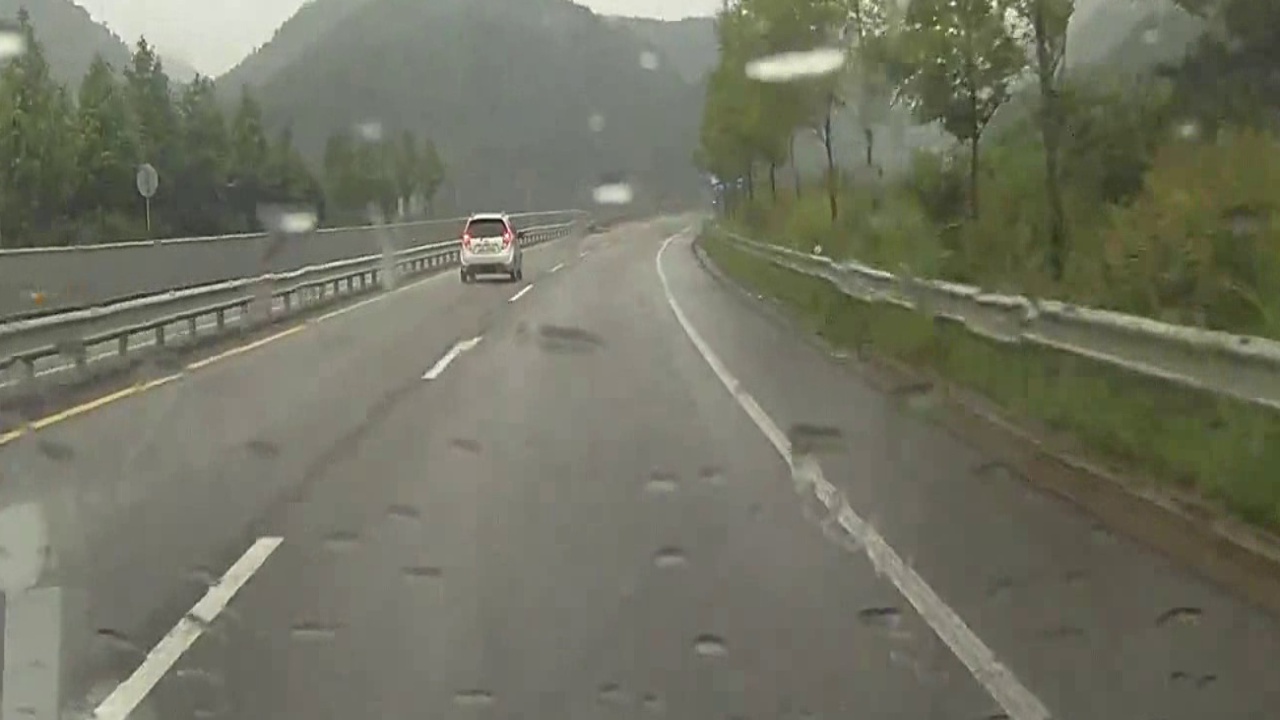}}
\hfil
\subfloat[]{\includegraphics[width=2in]{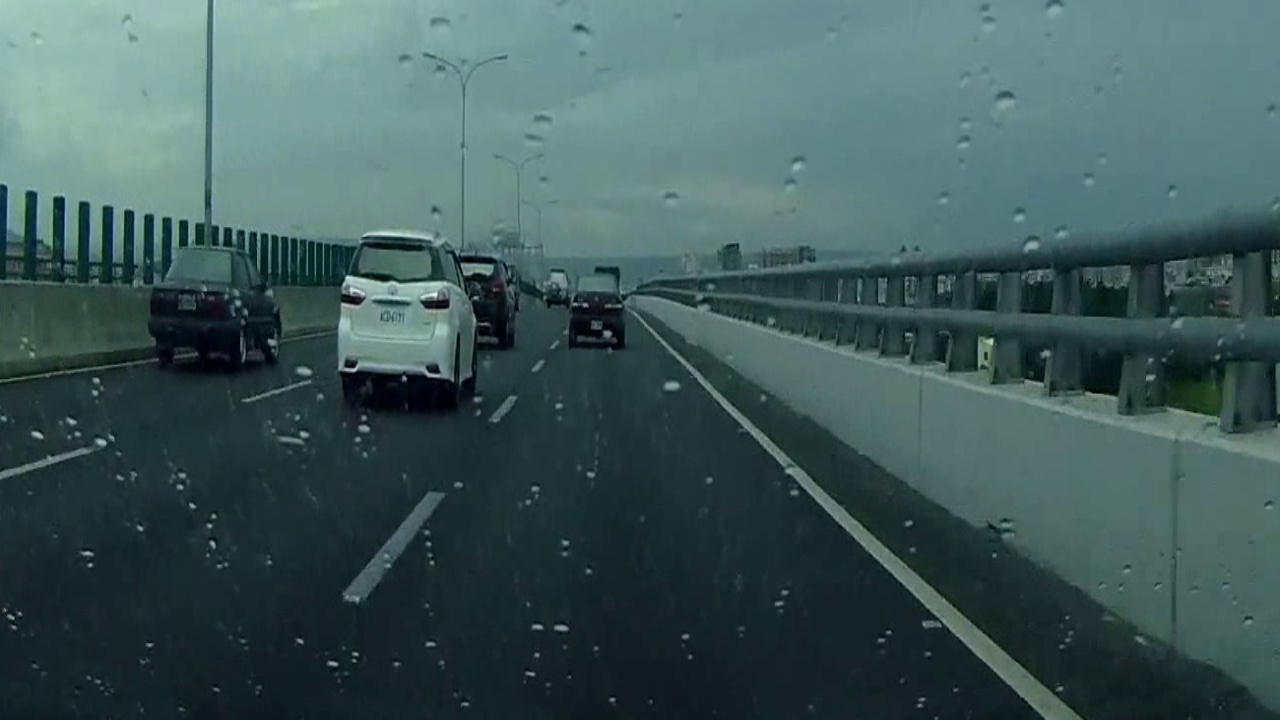}}
\hfil
\subfloat[]{\includegraphics[width=2in]{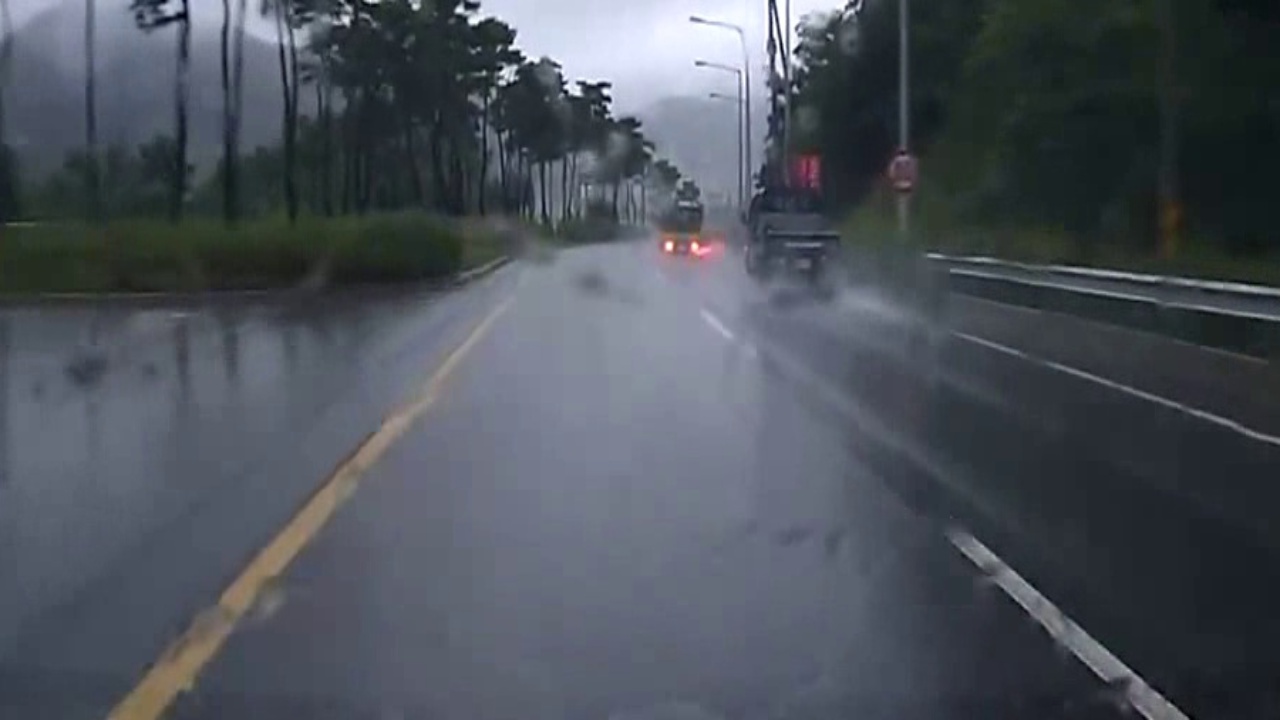}}
\hfil
\subfloat[]{\includegraphics[width=2in]{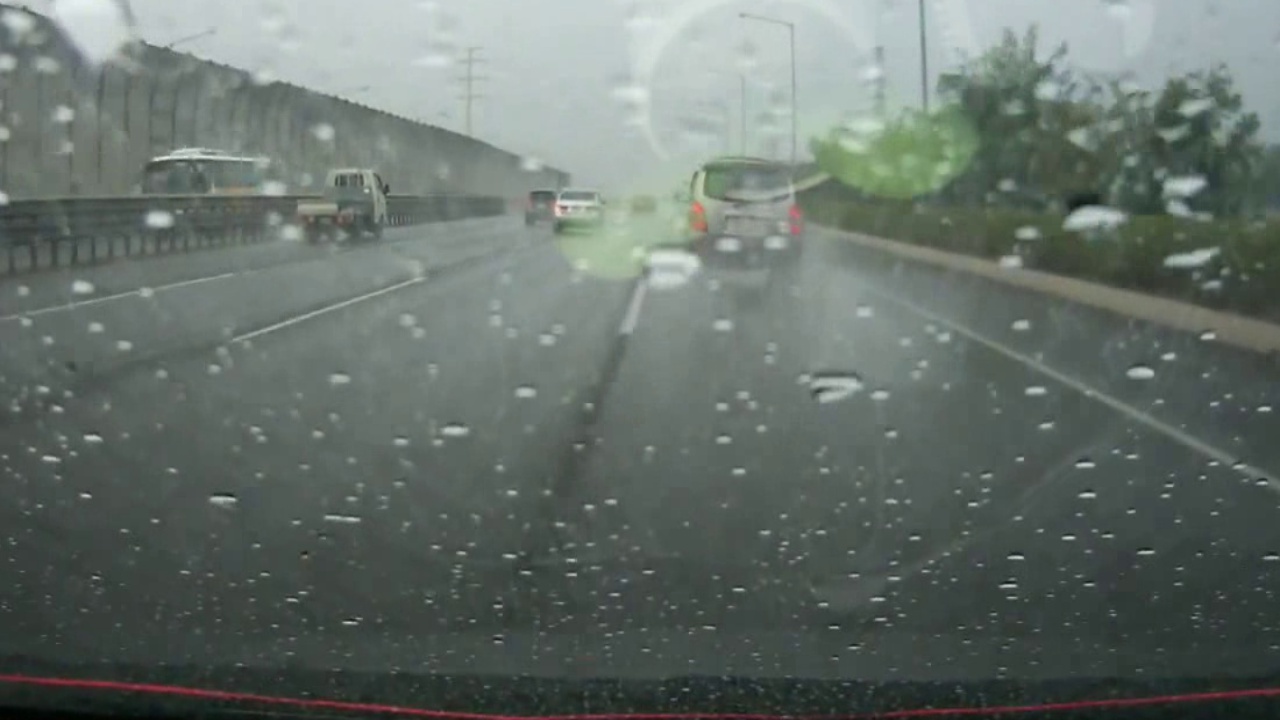}}
\hfil
\subfloat[]{\includegraphics[width=2in]{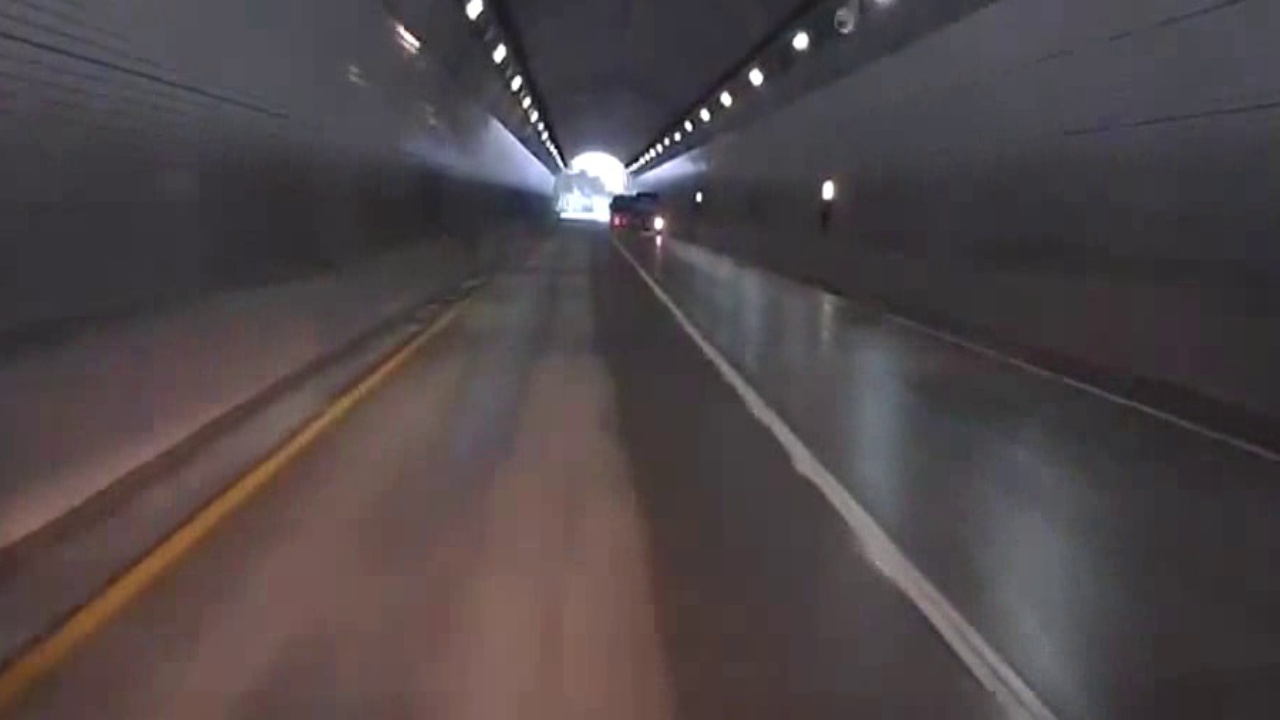}}
\caption{Example images of the DSDLDE dataset. (a), (b), and (c) are the mildly rainy cases while (d) and (e) are heavily rainy cases. (f) is the challenging cave scene included in the heavily rainy clips.}
\label{data_example}
\end{figure}

 
Rainy samples from the DSDLDE dataset were used to test the algorithms' adaptability toward different weather conditions. 5 video clips were selected from the ``DayRain'' category included in the DSDLDE dataset. All samples show visible raindrops on the windshield, significantly adding complexity to the lane detection task. According to the level that the lines were blurred by the raindrops, the 5 clips were further divided into ``Standard rainfall'' and ``Heavy rainfall'' categories. The example images of each category are shown in Figure. \ref{data_example}. The subfigures (a), (b), and (c) were captured from the ``Mildly rainy'' clips, while (d) and (e) were from the ``Heavily'' category. Notably, Figure. \ref{data_example} (f) shows a cave scenario captured from one of the Heavy rainfall samples. This scene added an additional challenge to the test dataset.

\subsection{Evaluation metrics}
Different from segmentation tasks, where IoU (Intersection over Union) and F-1 score are usually adopted, coordinates were used to define the lane marks and evaluate the detection results. To compare the results with previous research more effectively, the metrics of the Tusimple dataset were leveraged. All output from the CNN lane detection models was translated to the format of the ground truths provided by Tusimple and evaluated.
\par 
The metrics only considered the lane that the vehicle was in and the adjacent lanes on both sides. That is, the line quantity in the ground truths was capped at 4. Moreover, 56 y-coordinates were sampled on the images. The coordinates ranged from 160 to 710 with an increment of 10. Each detected line was expressed with an array of x-coordinates where it intersected with the above anchor coordinates. Each point in the detected line was compared with the x-coordinate of the ground truth. The accuracy of a line was derived by calculating the percentage of samples with a difference lower than a threshold value. The overall accuracy result was the average of all lines in the test dataset.

\section{Results}
\subsection{The ablation study for color channel selection}
In order to verify the selection of color channel allocation, an ablation study was performed on multiple models of different channel combinations. The results are shown in Table \ref{color_ablation}. According to the table, the original CLRNet showed excellent lane detection accuracy in clear weather conditions (Tusimple dataset). The colorful scenes during clear weather contained more information in all three color channels. When applying the proposed method, though the edge information contributed extra information, some information was lost due to the replacement of the RGB channels, leading to a slight accuracy decrease in some cases.
\par 
When replacing some channels with the processed edge information, an increase in accuracy was observed during rainy weather conditions. Though replacing all channels with the preprocessed information contributed to the highest accuracy (71.15\%) in heavy rain conditions, the model demonstrated inferior performance compared to that retained with the green channel. This is because the green channel contributed to most elements on the side of the roads, including grass, trees, bushes, and the yellow lane marks. Retaining the information in the green channel benefited the detection result in clear and mildly rainy weather. However, during heavy rain, the color contrast in images significantly decreased, making edge information the most reliable channel of all.
\par 
In order to achieve the best overall accuracy, the model that retained only the green channel was selected in this work.

\begin{table*}[b]
\caption{The ablation study among different color space combinations. (Unit:\%)}
\centering
\resizebox{\textwidth}{!}
{
\begin{tabular}{|c|ccc|ccc|}
\hline
\multirow{2}{*}{Model}           & \multicolumn{3}{c|}{Channel}                                                             & \multicolumn{3}{c|}{Datasets}                                                                     \\ \cline{2-7} 
                                 & \multicolumn{1}{c|}{0}             & \multicolumn{1}{c|}{1}              & 2             & \multicolumn{1}{c|}{Tusimple}       & \multicolumn{1}{c|}{DSDLDE(mild rain)} & DSDLDE(heavy rain) \\ \hline
CLRNet                           & \multicolumn{1}{c|}{\textcolor{blue}{Blue}}          & \multicolumn{1}{c|}{\textcolor{green}{Green}}          & \textcolor{red}{Red}           & \multicolumn{1}{c|}{96.63}          & \multicolumn{1}{c|}{84.34}             & 55.28              \\ \hline
\multirow{7}{*}{CLENet+Proposed} & \multicolumn{1}{c|}{Edge}          & \multicolumn{1}{c|}{Edge}           & Edge          & \multicolumn{1}{c|}{96.09}          & \multicolumn{1}{c|}{88.87}             & \textbf{71.15}              \\ \cline{2-7} 
                                 & \multicolumn{1}{c|}{Edge}          & \multicolumn{1}{c|}{\textcolor{green}{Green}}          & \textcolor{red}{Red}           & \multicolumn{1}{c|}{96.68}          & \multicolumn{1}{c|}{90.42}             & 62.37              \\ \cline{2-7} 
                                 & \multicolumn{1}{c|}{\textcolor{blue}{Blue}}          & \multicolumn{1}{c|}{Edge}           & \textcolor{red}{Red}           & \multicolumn{1}{c|}{96.69}          & \multicolumn{1}{c|}{92.93}             & 64.33              \\ \cline{2-7} 
                                 & \multicolumn{1}{c|}{\textcolor{blue}{Blue}}          & \multicolumn{1}{c|}{\textcolor{green}{Green}}          & Edge          & \multicolumn{1}{c|}{\textbf{96.74}}          & \multicolumn{1}{c|}{93.51}             & 66.64              \\ \cline{2-7} 
                                 & \multicolumn{1}{c|}{\textcolor{blue}{Blue}}          & \multicolumn{1}{c|}{Edge}           & Edge          & \multicolumn{1}{c|}{96.47}          & \multicolumn{1}{c|}{91.46}             & 66.65              \\ \cline{2-7} 
                                 & \multicolumn{1}{c|}{\textbf{Edge}} & \multicolumn{1}{c|}{\textbf{\textcolor{green}{Green}}} & \textbf{Edge} & \multicolumn{1}{c|}{96.63} & \multicolumn{1}{c|}{\textbf{94.22}}    & 66.78     \\ \cline{2-7} 
                                 & \multicolumn{1}{c|}{Edge}          & \multicolumn{1}{c|}{Edge}           & \textcolor{red}{Red}           & \multicolumn{1}{c|}{96.54}          & \multicolumn{1}{c|}{93.95}             & 65.65              \\ \hline
\end{tabular}
}

\label{color_ablation}
\end{table*}

\subsection{The accuracy and robustness of the model}
In order to verify the robustness of the proposed framework, the training process was repeated for 10 times and tested on both Tusimple and DSDLDE datasets. The results are shown in Table \ref{10times}. Compared to the original neural network (CLRNet), the preprocessing model had a higher average accuracy and robustness in all weather conditions. As discussed in the previous subsection, the advantage of the edge channel was not obvious during clear weather with abundant color information. Conversely, significant increases in accuracy were observed in dull, rainy cases.

\begin{table*}[t]
\caption{The repeated accuracy test of the proposed model}
\centering
\resizebox{0.9\textwidth}{!}{%
\begin{tabular}{|c|c|c|c|c|}
\hline
Weather                       & Algorithm       & Accuracy      & Precision  & Recall     \\ \hline
\multirow{2}{*}{Clear}        & CLRNet          & 96.63±0.14\%  & 0.98±0.003 & 0.97±0.004 \\ \cline{2-5} 
                              & CLRNet+Proposed & 96.63±0.04\%  & 0.98±0.002 & 0.97±0.001 \\ \hline
\multirow{2}{*}{Light   Rain} & CLRNet          & 84.34±17.17\% & 0.93±0.08  & 0.82±0.19  \\ \cline{2-5} 
                              & CLRNet+Proposed & 94.22±2.49\%  & 0.95±0.03  & 0.94±0.05  \\ \hline
\multirow{2}{*}{Heavy   Rain} & CLRNet          & 55.28±30.36\%   & 0.68±0.22  & 0.53±0.3   \\ \cline{2-5} 
                              & CLRNet+Proposed & 66.78±17.87\% & 0.81±0.11  & 0.62±0.13  \\ \hline
\end{tabular}
}
\label{10times}
\end{table*}

\subsection{The accuracy comparison with learning-based methods}
As mentioned in the literature review section, though a few hybrid lane detection methods have been proposed, the adaptability issue of these methods has not been addressed. As a result, incorporating the proposed module with state-of-the-art CNN algorithms and drawing comparisons with the original model demonstrates the improvement contributed by the module more effectively.
The proposed preprocessing algorithm was integrated with CLRNet, RESA, and Enet-SAD and tested on DSDLDE and tusimple datasets. The result is shown in Table. \ref{alldata}. And a demo video can be found at https://youtu.be/iC6yhfVLm9E
\par 
The proposed preprocessing algorithm was compatible with all three CNN-based lane detection models. By integrating the preprocessing algorithm, the detection performance of the models was significantly improved in rainy weather conditions despite a minor ($<1\%$) decrease in the clear daytime dataset. 
\par 
It was observed that the edge detection function in the proposed algorithm was able to overcome the challenges posed by the raindrops. A significant improvement was thus observed in the mildly rainy scenario. For example, the detection accuracy of the CLRNet was increased from 85.32\% to 95.00\% after applying the proposed preprocessing module. In addition, the algorithm showed its capability to improve the detection result of the seriously blurred images collected under heavy rainy scenarios as well. Since the models were only trained in Tusimple dataset, this result also highlighted the increased generalizability contributed by the proposed algorithm.

\begin{table}[t]
\centering
\caption{The accuracy of various models on the Tusimple dataset and the DSDLDE dataset.}
\resizebox{0.7\textwidth}{!}{%
\centering
\begin{tabular}{|c|c|c|c|}
\hline
\backslashbox{Model}{Dataset}                    & Tusimple & \begin{tabular}[c]{@{}c@{}}DSDLDE \\ Mild rain\end{tabular} & \begin{tabular}[c]{@{}c@{}}DSDLDE\\ Heavy rain\end{tabular} \\ \hline
CLRNet              & 96.78    & 85.32                                                  & 61.00                                                  \\ \hline
Proposed + CLRNet   & 96.64    & 95.00                                                  & 69.66                                                  \\ \hline
RESA                & 96.78    & 84.48                                                  & 58.93                                                  \\ \hline
Proposed + RESA     & 96.49    & 86.49                                                  & 59.09                                                  \\ \hline
Enet-SAD            & 83.30      & 71.20                                                  & 51.70                                                   \\ \hline
Proposed + Enet-SAD & 91.50     & 88.55                                                  & 59.73                                                  \\ \hline
\end{tabular}%
}

\label{alldata}
\end{table}

\subsection{The accuracy comparison with geometric-based methods}
The performance of the proposed algorithm was also compared with geometric-based methods mentioned in the previous section. Both research\citep{ashape}\citep{rain_threshold} considered only the closest two lines. Since the evaluation of \citep{rain_threshold} was conducted manually, the evaluation metrics of \citep{ashape} were adopted in the comparison. According to Lee et al\citep{ashape}, the detected lines that overlap the lower 2/3 of the ground truth are defined as ``detected''. When switching lanes, the allowed overlap region was increased by $50\%$. However, whether the vehicle is switching lanes cannot be clearly defined. The proposed algorithm was tested with the original standard on all images.
\par 
\begin{table}[t!]
\centering
\caption{The comparison with geometric-based methods on the DSDLDE dataset}
\begin{tabular}{|p{0.3\textwidth}|p{0.15\textwidth}|m{0.35\textwidth}|}
\hline
Models            & Accuracy(\%) & \multicolumn{1}{c|}{Additional notes}  \\ \hline
CLRNet            & 84.86    &                                        \\ \hline
Proposed + CLRNet & 95.81    &                                        \\ \hline
Lee\citep{ashape}               & 96.9     & Lowered standard while switching lanes \\ \hline
Ahmed\citep{rain_threshold}             & 92-97    & Disregard switching lane images        \\ \hline
\end{tabular}
\label{geometric}
\end{table}

The comparison result is shown in Table. \ref{geometric}. All models were tested on the ``DayRain'' images from the DSDLDE dataset. The result shows that the accuracy of the proposed algorithm is similar to the previous works under a more difficult standard. In addition, the proposed algorithm was able to conduct multiple lane and curved lane detection tasks while geometric-based methods were limited to single straight lanes.

\subsection{Processing time}
The speed of the preprocessing algorithm was measured using the machine mentioned above. Since the preprocessing algorithm was an additional module applied to CNNs, there was an unavoidable time increase. The processing time required for CLRNet was 21.5 ms, while that with the preprocessing model was 30.2 ms. The overall processing speed still falls in the real-time frame for 30fps cameras. The additional processing time contributed to an increased robustness/accuracy. With a 9 ms increase in the processing time, the system is able to gain 11.7/20.8\% of accuracy in light/heavy rain weather conditions.

\section{Conclusion}
An innovative hybrid approach to lane detection was proposed in the work. The experimental result showed that the proposed framework was able to integrate with various CNN-based lane detection models. The fuzzy parameter tuner provided a more informative input for the neural networks and, therefore, enhanced the generalizability and accuracy during challenging weather conditions. 
\par 
There are some limitations in the proposed algorithm. As mentioned above, since the module was an addition to the neural network, an unavoidable increase in the processing time was observed. Due to the use of neural networks, onboard computers supported by GPU capability are required when applying the proposed algorithm to real world systems. In addition, it required several frames for the system to stabilize before reaching an optimized parameter because the values were tuned gradually with each incoming frame. When encountering scenes with very fast illumination changes, adjustments need to be made to minimize the reaction time.
\par 
More variations of hybrid image processing algorithms can be expected in future research. The fuzzy logic-based parameter tuner also has the potential to integrate with other geometric-based algorithms that require manual tuning. Furthermore, different approaches can be developed in the color space allocation function. For example, the data from certain channels can be further enhanced and processed to retrieve additional information. A similar channel allocation approach could also be applied to other vision-based tasks such as drivable region detection, obstacle detection, and traffic sign recognition. Finally, implementation of this technology within an autonomous vehicle platform constitutes a pivotal next phase.

\section*{Acknowledgements}

The authors would like to thank TechnipFMC plc for the gift grant used in funding this research through the University of Illinois Urbana-Champaign.



\bibliographystyle{apacite} 
\biboptions{authoryear}
\bibliography{refs}





\end{document}